\documentclass{iau}
\usepackage{graphicx}
\usepackage{amsmath}

\title[Mode selection in pulsating stars] 
{Mode selection in pulsating stars}

\author[Rados\l{}aw Smolec]   
{Rados\l{}aw Smolec}

\affiliation{Nicolaus Copernicus Astronomical Centre, \\ ul. Bartycka 18, 00-716 Warszawa, Poland \\ email: {\tt smolec@camk.edu.pl}}

\pubyear{2013}
\volume{301}  
\pagerange{119--126}
\jname{Precision Asteroseismology}
\editors{W. Chaplin, J. Guzik, G. Handler \& A. Pigulski, eds.}
\begin{document}

\maketitle

\begin{abstract}
In this review we focus on non-linear phenomena in pulsating stars the mode selection and amplitude limitation. Of many linearly excited modes only a fraction is detected in pulsating stars. Which of them and why (the problem of mode selection) and to what amplitude (the problem of amplitude limitation) are intrinsically non-linear and still unsolved problems. Tools for studying these problems are briefly discussed and our understanding of mode selection and amplitude limitation in selected groups of self-excited pulsators is presented. Focus is put on classical pulsators (Cepheids and RR Lyrae stars) and main sequence variables ($\delta$~Scuti and $\beta$~Cephei stars). Directions of future studies are briefly discussed. 
\end{abstract}

\firstsection 
\section{Introduction}

The models of pulsating stars typically predict more unstable modes than is observed. Which of the linearly unstable modes are excited and why -- the problem of mode selection -- is a difficult non-linear problem, still lacking a satisfactory solution. Closely related is a problem of amplitude limitation, which is non-linear as well. Intrinsic non-linearity of these two problems is a major challenge. Our tools to analyse the non-linear pulsation are either restricted to large amplitude radial pulsation, e.g. in Cepheids and RR~Lyrae stars (hydrodynamical modelling) or are based on simplified assumptions and depend on the unknown parameters (amplitude equations formalism). Therefore these problems received only scant theoretical attention in the past and dated but excellent review of \cite[Dziembowski (1993)]{D93} is still mostly up-to-date. 

For the ground-based observations the basic mode selection mechanism is of observational nature. Because of geometric cancellation  modes of degree $\ell>2$ are hard to detect from the ground (\cite[Dziembowski 1977]{D77}). In the space-based photometry mode degrees above $10$ are reported (e.g. \cite[Poretti et al. 2009]{poretti}) and since geometrical cancellation is very similar for large $\ell$ it is hard to point any obvious limit for maximum $\ell$, except that even-$\ell$ modes are less affected by cancellation and thus more likely to be detected. In this review we focus on intrinsic non-linear mechanisms acting in pulsating stars.
  
In the next Section the tools for studying the non-linear phenomena are briefly discussed. Then we discuss the mode selection mechanisms: the mode trapping (Section~\ref{sec.trapping}), non-resonant and resonant mode interaction (Section~\ref{sec.coupling}) and next turn to the discussion of amplitude limiting effects: collective saturation of the driving mechanism (Section~\ref{sec.collective}) and resonant mode coupling (Section~\ref{sec.dsct}). Discussion and outlook for the future studies end this review.

\section{Tools for mode selection analysis}\label{sec.tools}

\begin{figure}[t]
\begin{center}
\includegraphics[width=5in]{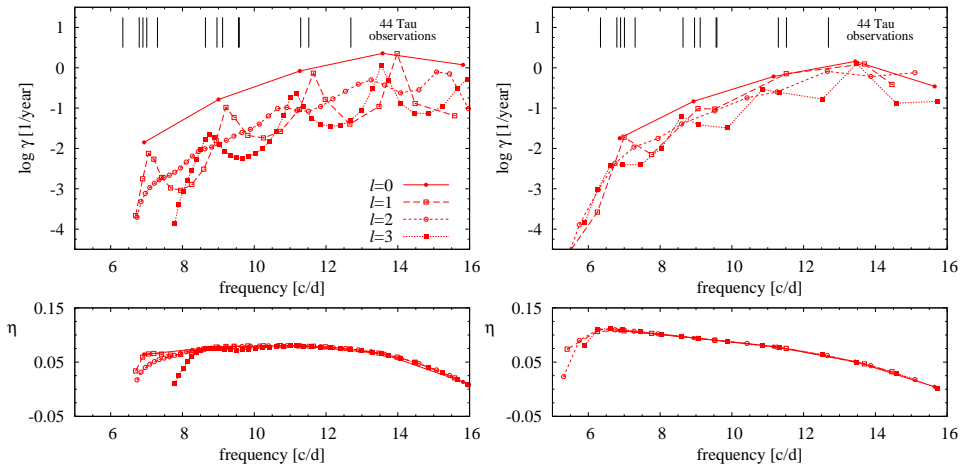} 
\end{center}
\caption{Linear growth rates, $\gamma$ (top) and Stellingwerf's growth rates, $\eta$ (bottom) for two models of $\delta$~Sct stars at different evolutionary stages: post-main sequence (MS) expansion (left) and post-MS contraction phase (right). In the top panels frequencies detected in 44~Tau are marked. Models from \cite[Lenz et al. (2008, 2010)]{Lenz08,Lenz10}.}
\label{fig1}
\end{figure}

{\underline{\it Linear stability analysis.}} The  linear stability analysis tells nothing about the mode selection or amplitudes of the excited modes -- these are non-linear problems. It is, however, a necessary starting point, as it provides the information on mode eigenfunctions, mode frequencies, $\sigma$, and on mode stability through the linear growth rate, $\gamma$:
\begin{equation}\gamma=\frac{\int dW}{2\sigma I}\,,\end{equation}
where
\begin{equation}dW=\Im\bigg[\delta P\bigg(\frac{\delta\rho}{\rho}\bigg)^*\bigg]\,,\,\,\,\, I=\int_M |\boldsymbol{\xi}|^2dm\,,\end{equation}
are local contribution to the work integral ($dW$, with pressure, $\delta P$, and density, $\delta\rho$, perturbations) and mode inertia ($I$, with radial displacement, $\boldsymbol{\xi}$). Plots of growth rates for low degree modes in a $\delta$~Sct type models are shown in the top panels of Fig.\,\ref{fig1}. The growth rates of non-radial modes are not smooth, but exhibit maxima, particularly pronounced for the evolved model (left panel) and modes of $\ell\!=\!1$. This peculiar frequency dependence of the growth rates reflects the behaviour of mode inertia. Modes trapped in the external layers of the model with small amplitudes in the interior have the smallest inertia and the highest growth rates. The inference that the most unstable, trapped modes will be most easily driven to high amplitude is precarious, however (see Section~\ref{sec.trapping}). The maxima of the growth rates do not reflect the properties of the driving region, which is clear if the \cite[Stellingwerf's (1978)]{Stel78} growth rates are considered instead:
\begin{equation}\eta=\frac{\int dW}{\int |dW|}\,,\end{equation}
\noindent with $\eta\in(-1,1)$. These growth rates are plotted in the bottom panels of Fig.\,\ref{fig1} and they smoothly vary with the mode frequency.

{\underline{\it Hydrodynamic models.}} Realistic non-linear hydrodynamic models of radially pulsating stars are computed for nearly 50 years now (e.g. \cite[Christy 1966]{christy}). Computations are done with direct time-integration, one dimensional codes. The initial static structure is perturbed with the scaled velocity eigenfunction (initial {\it kick}) and time evolution of the model is followed till the finite amplitude steady pulsations are reached (limit cycle). As different initializations may lead to different limit cycles for the same static model (hysteresis), mode selection analysis is time consuming and requires computation of tenths of models. The convergence may be speed-up with the use of relaxation technique (\cite[Stellingwerf 1974]{Stel74}), which in addition provides the information about stability of the limit cycles (even unstable ones) through the Floquet exponents.

 The early codes were purely radiative. Currently several codes that include convective energy transfer are in use. Two prescriptions for the turbulent convection are commonly adopted, either by \cite[Stellingwerf (1982)]{Stel82} (e.g. in the Italian code, \cite[Bono \& Stellingwerf 1992]{bs92}) or by \cite[Kuhfu\ss{} (1986)]{kuhfuss} (e.g. in the Warsaw codes of \cite[Smolec \& Moskalik (2008a)]{SM08a} or in the Florida-Budapest code, e.g. \cite[Koll\'ath et al. (2002)]{kollath02}, with the modified Kuhfu\ss{} model). Both models include several free parameters, values of which must be adjusted to match the observational constraints. 

Non-linear pulsation codes were successfully used to model the light and radial velocity curves in single-periodic classical pulsators. Understanding of the dynamical phenomena shaping these curves would not be possible however without the insight provided by the analysis of amplitude equations. 

{\underline{\it Amplitude equations (AEs).}} If the growth rates of the dominant modes are small compared to their frequencies (weak non-adiabaticity), and assuming weak non-linearity, the hydrodynamic equations governing the stellar pulsation may be reduced to ordinary differential equations for the amplitudes of the excited modes, $A_i$ (e.g. \cite[Dziembowski 1982]{D82}, \cite[Buchler \& Goupil 1984]{bg84}). In case no resonances are present among pulsation modes the form of the AEs (usually truncated at the cubic terms) is the following:
\begin{equation}\frac{dA_i}{dt}=\gamma_i\Big(1+\sum_j\alpha_{ij}A_j^2\Big)A_i\,,\end{equation}
where $\alpha_{ii}$ and $\alpha_{ij}$ are negative self- and cross-saturation coefficients, respectively.

In case of resonant mode coupling the exact form of amplitude equations depends on the resonance considered. Here we present the complex equations for the parametric resonance $\sigma_a=\sigma_b+\sigma_c+\Delta\sigma$:
\begin{align}
\frac{dA_a}{dt}&=\gamma_aA_a-i\frac{C}{2}A_bA_c{\rm e}^{-i\Delta\sigma t}\,,\nonumber\\
\frac{dA_{b,c}}{dt}&=\gamma_{b,c}A_{b,c}-i\frac{C}{2}A_aA_{c,b}^*{\rm e}^{i\Delta\sigma t}\,.
\end{align}
$C$ is a resonant coupling coefficient.

With reasonable approximations the amplitude equations may be solved analytically. Of particular interest are time-independent solutions (fixed points) which correspond to limit cycles in hydrodynamic computations. Analysis of fixed point's stability provides a direct insight into mode selection. For non-resonant AEs the single-mode fixed points are given by $A_i=1/\sqrt{-\alpha_{ii}}$ and are stable if $\alpha_{ji}/\alpha_{ii}>1$ for each $j$. For the interesting case of non-resonant two mode interaction the double-mode solution, with finite amplitude of the two excited modes is possible once $\alpha_{00}\alpha_{11}-\alpha_{01}\alpha_{10}>0$ (analysis of cubic AEs), i.e. when the self-saturation exceeds the cross-saturation. The detailed discussion of mode selection scenarios for both non-resonant and resonant mode interaction may be found e.g. in \cite[Dziembowski \& Kov\'acs (1984)]{DKov84} or \cite[Buchler \& Kov\'acs (1986)]{BK86}.

The described mode selection analysis is possible only when the values of the saturation/coupling coefficients are known. These however, are very difficult to compute. Only with simplistic approximations some analytical estimates are possible. Therefore, most of the work on AEs are parametric studies. This problem may be overcome for large-amplitude radial pulsators for which hydrodynamic computations may be coupled with the analysis of AEs. For time integration of the same model, but initialized with different initial conditions, the evolution of mode amplitudes may be followed with the help of analytical signal method (e.g. \cite[Koll\'ath et al. 2002]{kollath02}). The resulting trajectories are then fitted with the appropriate AEs and resulting saturation/coupling coefficients may be used to compute all the fixed points and their stability, i.e. to analyse the mode selection. Repeating the procedure for a discrete set of models located at different parts of the HR diagram, and interpolating in-between, yields a consistent picture of mode selection in the full instability strip (\cite[Szab\'o, Koll\'ath \& Buchler 2004]{SKB04}). Results of such analysis for Cepheids are reported in Section~\ref{sec.coupling}.

\section{Mode trapping}\label{sec.trapping}

The mode trapping as a mode selection mechanism was first proposed by \cite[Winget, Van Horn \& Hansen (1981)]{winget81} in the context of white dwarf (ZZ~Ceti) pulsations. Trapping in the strongly stratified models of white dwarfs is caused by the resonance between the wavelength of the g-mode and the thickness of one of the compositional layers. The trapped modes have low amplitude in the core, with most of the mode energy confined in the outer regions, where pulsation driving takes place. The mode inertia is low and the growth rate is high (eq. 2.1). Winget et al. concluded that the trapped modes are much more likely to be excited than adjacent, non-trapped modes. The mode trapping is also present in the models of evolved $\delta$~Sct stars, as is clearly visible in Fig.\,\ref{fig1} (top, left). In this case, the frequency separation between the trapped modes corresponds to the separation between the consecutive radial overtones. \cite[Dziembowski \& Kr\'olikowska (1990)]{DK90} proposed that mode trapping might be a mode selection mechanism in the evolved $\delta$~Sct stars. They comment however, that this selection rule relies only on linear non-adiabatic theory and since the high mode growth rates are not indicators of large amplitude, justification must come from the non-linear theory. 

\begin{figure}[t]
\begin{center}
\includegraphics[width=4.5in]{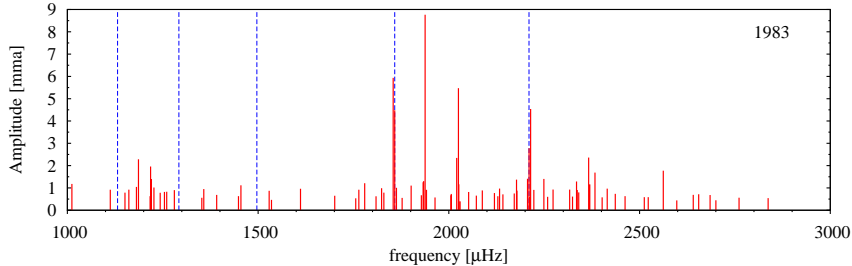} 
\end{center}
\caption{Frequency spectrum of PG\thinspace 1159-035 (\cite[Costa et al. 2008]{costa}) from the 1983 season. Frequencies of trapped modes are marked with dashed lines.}
\label{fig2}
\end{figure}

The observations itself invalidate the mode trapping as a mode selection rule. In case of white dwarfs the mode trapping is clearly detected through characteristic wave shape of the period spacing vs. period diagram, which allows the identification of the trapped modes. It turns, that these are not the trapped modes that have the highest amplitudes. As an example we use the observations of PG 1159-035 -- a hot pulsator, but evolved and stratified enough to show the effects of mode trapping (\cite[Costa et al. 2008]{costa}). In Fig.\,\ref{fig2} we show the frequency spectrum of the star from the 1983 season. The dashed lines mark the location of the trapped modes derived from the period spacing diagrams constructed using six seasons of observations. Although the two highest frequency trapped modes have high amplitudes, the neighbouring non-trapped mode has the highest amplitude. For the three low frequency trapped modes no signal was detected in 1983 season. For two of these modes significant detection was made only during one out of the six observing seasons. 

In case of $\delta$~Sct stars regularities observed in the frequency spectra are also interpreted as a result of mode trapping. \cite[Breger, Lenz \& Pamyatnykh (2009)]{blp09} show that in the case of FG~Vir, BL~Cam and 44~Tau there is a preferred frequency spacing between the excited modes, which corresponds to the spacing between radial modes. The asteroseismic model of \cite[Lenz et al. (2008)]{Lenz08} pointed that mode trapping may be indeed operational in 44~Tau. The growth rates for their best asteroseismic model, located in the post-MS expansion phase are reproduced in Fig.\,\ref{fig1} (top left). Observed modes seem to cluster around the growth rate maxima. However, this model fails to reproduce all the observable parameters satisfactorily. In their later analysis, \cite[Lenz et al. (2010)]{Lenz10} constructed asteroseismic models at the earlier evolutionary phase, the post-MS contraction, and obtained a better model with an excellent fit of all the observed modes (Fig.\,\ref{fig1}, top right). The mode trapping is only barely marked for this model and cannot represent a valid mode selection mechanism.

\section{Resonant and non-resonant mode coupling}\label{sec.coupling}

The analysis of mode selection and amplitude limitation is most feasible for large amplitude, radial pulsators, Cepheids and RR Lyrae stars, as direct hydrodynamic models may be computed and complementary analysis of amplitude equations is feasible. Most of these stars are single-periodic, pulsating either in the fundamental (F) mode or in the first overtone (1O). In many stars double-mode pulsation either of the F+1O or 1O+2O type is detected (see \cite[Moskalik 2013]{pam13} and these proceedings for a review). Already with first, purely radiative hydrocode, \cite[Christy (1966)]{christy} showed that in the single-periodic models amplitude growth is limited by the saturation of the driving mechanism. The selection between the fundamental and first overtone pulsation is however a still unsolved problem. The analysis of radiative models, yielded the following picture (e.g. \cite[Stellingwerf 1975]{Stel75}). If only one mode is linearly unstable, then this mode reaches the finite amplitude: first overtone at the blue side of the instability strip and fundamental mode at the red side. If two modes are simultaneously unstable, then either only one limit cycle is stable (F-only or 1O-only domains) or two limit cycles are simultaneously stable, and which is selected depends on the initial conditions (E/O, either-or domain). The star entering the E/O domain from the blue side will continue to pulsate in the 1O mode, while star entering from the red side will continue to pulsate in the F mode. No double-mode pulsation was found in realistic radiative models of Cepheids and RR Lyrae stars.

Inclusion of turbulent convection into the models seemed to solve the problem. \cite[Feuch\-tin\-ger (1998)]{feuchtinger} reported one double-mode RR~Lyrae model and Florida-Budapest group found the double-mode Cepheids and RR~Lyrae models in their surveys (\cite[Koll\'ath et al. 1998, 2002]{kollath98,kollath02}, \cite[Szab\'o, Koll\'ath \& Buchler 2004]{SKB04}, \cite[Buchler 2009]{buchler09}). Regrettably, how the inclusion of turbulent convection caused the stable double-mode pulsation in these models was not analysed. In \cite[Smolec \& Moskalik (2008b)]{SM08b} we were able to show that the double-mode pulsation was caused by unphysical neglect of buoyant forces in convectively stable regions of the model. At the absence of restoring force, the turbulence is not damped effectively below the envelope convective zone and resulting strong eddy-viscous dissipation reduces the amplitudes of fundamental and first overtone modes differentially, favouring the occurrence of double-mode pulsation. With the correct treatment of the buoyant forces we were not able to find satisfactory double-mode Cepheid models. Also the computations with Italian code, adopting the Stellingwerf's model of convection yielded null result (see \cite[Smolec \& Moskalik 2010]{SM10}).

Although the resonant mode interaction cannot explain the double-mode pulsation for most of the observed variables it may be operational in some limited parameter ranges. In particular the 2:1 resonance between the fundamental mode and the linearly damped second overtone may decrease the amplitude of the former mode, allowing the growth of the first overtone, as pointed by \cite[Dziembowski \& Kov\'acs (1984)]{DKov84}. Some hydrodynamic resonant double-mode models where in fact found (\cite[Smolec 2009]{S09}, \cite[Buchler 2009]{buchler09}) and the two long period double-periodic Cepheids discovered recently in M31 by \cite[Poleski (2013)]{poleski} are first good candidates for resonant double-periodic pulsation. For the majority of double-periodic pulsators explanation is still missing.

\section{Collective saturation of the driving mechanism}\label{sec.collective}

Among $\delta$~Sct stars and $\beta$~Cep stars there are variables with one or two dominant radial modes, with amplitudes of order of 0.1\thinspace mag. The attempts to model these stars with hydrodynamic codes failed however. \cite[Stellingwerf (1980)]{Stel80} computed $\delta$~Sct models and got pulsation amplitudes exceeding 1\thinspace mag (which he called {\it main sequence catastrophe}). Clearly, instability cannot be saturated with single pulsation mode in these stars, as is the case for Cepheids or RR~Lyr stars, occupying the high luminosity part of the same instability strip. Similar results were obtained for models of singly-periodic $\beta$~Cep pulsators computed by \cite[Smolec \& Moskalik (2007)]{SM07}. Linear stability computations predict that many non-radial acoustic modes are unstable in these stars. Smolec \& Moskalik assumed that the instability is collectively saturated not by a single mode, but by tenths ($n$) of acoustic modes simultaneously (and because of the assumed large $\ell$ these modes are not detected). Using amplitude equations and assuming that the properties of acoustic modes (saturation coefficients) are the same, one may show that in this case the amplitude drops by a factor $\sqrt{n}$ as compared to single-mode saturation amplitude predicted by hydrodynamic model. The agreement with pulsation amplitudes of multi-periodic $\beta$~Cep stars is obtained with only a fraction of the available (linearly unstable) modes. In principle, the collective saturation of the instability mechanism explains the amplitudes of $\beta$~Cep stars (and $\delta$~Sct stars as well), however, there is a serious difficulty with such explanation. The resulting macroturbulence velocities (and hence the line-widths) are too high as compared with observations, pointing that other amplitude limiting mechanism must be operational.

\section{Amplitude limitation in $\delta$~Scuti stars}\label{sec.dsct}

An alternative scenario to collective saturation of the driving mechanism is resonant mode coupling investigated by \cite[Dziembowski (1982)]{D82}, who considered the coupling of the unstable acoustic mode to a pair of stable g-modes. The AEs appropriate for this case were given in Section~\ref{sec.tools}. The parametric excitation of the g-modes start once the amplitude of the acoustic mode exceeds a critical value, $A_a\!>\!A_{\rm crit}$. Steady state solution is then possible in a limited range of mismatch parameter, and provided that damping of the gravity-mode pair exceeds the acoustic mode driving. The amplitude of the acoustic mode is then close to the critical amplitude, while amplitudes of the g-modes are much lower, making their detection from the ground impossible. The exact formulae for stability condition and critical/equilibrium amplitudes may be found in \cite[Dziembowski (1982)]{D82}. \cite[Dziembowski \& Kr\'olikowska (1985)]{DK85} applied the formalism for realistic $\delta$~Sct models. Strong coupling arises only if the radial orders of the gravity modes are close which implies $\sigma_b\!\approx\!\sigma_c\!\approx\!\sigma_a/2$ and for close and large $\ell$ values. Because of a large number of potential resonant pairs \cite[Dziembowski \& Kr\'olikowska (1985)]{DK85} computed the probability distribution for the critical amplitude. Their results pointed that indeed, the resonant mode coupling may be a promising amplitude limitation mechanism. The typical critical amplitude they found is of order of $0.01$\thinspace mag. Moreover, if rotation is taken into account the critical amplitude drops even further, as denser g-mode spectrum allows for fine tuning of the resonance condition (\cite[Dziembowski, Kr\'olikowska \& Kosovitschev 1988]{DKK88}). Thus, resonant mode coupling nicely explains the observational fact that high amplitude $\delta$~Sct stars are slow rotators.

The resonant mode coupling scenario has serious shortcomings, however. Only for low order acoustic modes, which couple to strongly damped global g-modes there is a large probability that equilibrium is stable. Higher order p-modes couple preferentially to weakly damped inner g-modes which are not able to halt the amplitude growth. The excitation of many g-mode pairs is then expected, and has to be analysed numerically, which was done by \cite[Nowakowski (2005)]{N05}. His results are disappointing however. Static multi-mode solution is not possible then, and strong amplitude variability on the $\gamma_a^{-1}$ time-scale is expected. Moreover, computations for realistic model of evolved $\delta$~Sct star XX Pyx show, that resonant mode coupling cannot be a dominant amplitude limiting effect as predicted amplitudes are higher than observed. Whether saturation of the driving mechanism plays a role in evolved $\delta$~Sct models was not investigated in detail yet.

\section{Discussion and conclusions}

The problems of mode selection and amplitude limitation are one of the most stubborn, still unsolved problems of stellar pulsation theory. They are important for all groups of pulsating stars and for none we have a satisfactory solutions. Even for large amplitude radial pulsators, we do not understand the mechanisms behind the simplest form of multi-mode pulsation, ie. double-mode pulsation. Triple-mode pulsation and excitation of non-radial modes are even more challenging problems. It seems that their solution must await the development of full 3D hydrodynamic models. Fortunately such codes are now being developed (\cite[Geroux \& Deupree 2013]{GD13}, \cite[Mundprecht, Muthsam \& Kupka 2012]{MMK12}), but were not applied for modelling the double-periodic pulsation yet.

For low-amplitude non-radial pulsators our understanding is even poorer, as use of the amplitude equations formalism, the only available tool to study the non-linear and non-radial pulsation, is strongly limited by their complexity and unknown saturation/coupling coefficients.

The most interesting quantities observations can provide are intrinsic amplitudes of pulsation modes. For their determination the mode identification and inclination degree are needed. The robust determination of these quantities is however difficult and requires combination of multi-band photometric and spectroscopic observations (Uytterhoeven, these proceedings). Only for limited number of main sequence pulsators and usually only for a few detected modes robust mode identifications are available. In case of stars studied with space telescopes with hundreds of detected modes the task seems even more challenging, if possible at all. The space observations, in particular their statistical analysis are however of great importance for our understanding of mode selection (\cite[Balona \& Dziembowski 1999]{BD99}). The frequency distribution of amplitudes of the excited modes may be used to infer the probability distribution of intrinsic amplitudes, assuming some knowledge on the $\ell$ values, and then compared with model computations. {\it Kepler} observations of thousands of $\delta$~Sct stars (\cite[Balona \& Dziembowski 2011]{BD11}) make such analysis feasible.  

We stress the need for systematic spectroscopic observations of targets of current and future space missions. The aim is not only the mode identification, but also precise determination of basic stellar parameters, $\log g$ and $\log T_{\rm eff}$, which are additional important constraints of seismic models and are necessary to study the intriguing problem of significant contamination of the $\delta$~Sct instability strip with apparently non-pulsating stars (\cite[Balona \& Dziembowski 2011]{BD11}).

{\bf Acknowledgements.} I am grateful to Wojtek Dziembowski, Pawe\l{} Moskalik and Alosha Pamyatnykh for many fruitful discussions and to Patrick Lenz for providing data for Fig.\,\ref{fig1}. I acknowledge the IAU grant for the conference. This research is supported by the Polish National Science Centre through grant DEC-2012/05/B/ST9/03932.

\end{document}